\documentclass[aps,pra,twocolumn,showpacs,superscriptaddress,floatfix]{revtex4-1}
\usepackage{graphicx,amsmath,amssymb}
\usepackage[usenames]{color}
\usepackage[dvipsnames]{xcolor}
\usepackage[unicode=true,pdfusetitle,
 bookmarks=false,
 breaklinks=false,pdfborder={0 0 1},backref=false,colorlinks,urlcolor=Black, linkcolor=blue,citecolor=blue]
 {hyperref}
\begin{document}

\title{
Quantum Fisher information and polaron picture for identification of transition coupling in quantum Rabi model
}
\author{Zu-Jian Ying}
\email{yingzj@lzu.edu.cn}
\affiliation{School of Physical Science and Technology, Lanzhou University, Lanzhou 730000, China}
\affiliation{Key Laboratory for Quantum Theory and Applications of MoE, Lanzhou Center for Theoretical Physics, and Key Laboratory of Theoretical Physics of Gansu Province, Lanzhou University, Lanzhou, Gansu 730000, China}

\author{Wen-Long Wang}
\affiliation{School of Physical Science and Technology, Lanzhou University, Lanzhou 730000, China}
\affiliation{Key Laboratory for Quantum Theory and Applications of MoE, Lanzhou Center for Theoretical Physics, and Key Laboratory of Theoretical Physics of Gansu Province, Lanzhou University, Lanzhou, Gansu 730000, China}

\author{Bo-Jian Li}
\affiliation{School of Physical Science and Technology, Lanzhou University, Lanzhou 730000, China}
\affiliation{Key Laboratory for Quantum Theory and Applications of MoE, Lanzhou Center for Theoretical Physics, and Key Laboratory of Theoretical Physics of Gansu Province, Lanzhou University, Lanzhou, Gansu 730000, China}

\begin{abstract}
The quantum Rabi model (QRM) is a fundamental model for light-matter interactions. A fascinating feature of the QRM is that it manifests a quantum phase transition which is applicable for critical quantum metrology (CQM). Effective application for CQM needs the exact location of the transition point, however the conventional expression for the transition coupling is only valid in the extreme limit of low frequency, while apart from zero frequency an accurate location is still lacking. In the present work we conversely use the quantum Fisher information (QFI) in the CQM to identify the transition coupling, which finds out that transition coupling indeed much deviates from the conventional one once a finite frequency is turned on. Polaron picture is applied to analytically reproduce the numeric QFI. An accurate expression for the transition coupling is obtained by the inspiration from the fractional-power-law effect of polaron frequency renormalization. From the QFI in the polaron picture we find that the transition occurs around a point where the effective velocity and the susceptibility of the single-photon absorption rate reach maximum. Our result provides an accurate reference of transition couplings for quantum metrology at non-zero frequencies. The formulation of the QFI in the polaron picture also prepares an analytic method with an accurate compensation for the parameter regime difficult to access for the numerics. Besides the integer/fractional power law analysis to extract the underlying physics of transition, the QFI/velocity relation may also add some insight in bridging the QFI and transition observables.
\end{abstract}
\pacs{ }
\maketitle


\section{Introduction}

In the modern research trends of light-mamter intercations induced by both
theoretical \cite{Braak2011,Solano2011,Boite2020,Liu2021AQT} and
experimental \cite{Diaz2019RevModPhy,Kockum2019NRP} progresses, few-body quantum phase transitions~\cite%
{Liu2021AQT,Ashhab2013,Ying2015,Hwang2015PRL,Ying2020-nonlinear-bias,Ying-2021-AQT,LiuM2017PRL,Hwang2016PRL,Irish2017, Ying-gapped-top,Ying-Stark-top,Ying-Spin-Winding,Ying-2018-arxiv,Ying-Spin-Winding,Ying-JC-winding,Grimaudo2022q2QPT,Grimaudo2023-Entropy}
and topological phase transitions~\cite%
{Ying-2021-AQT,Ying-gapped-top,Ying-Stark-top,Ying-Spin-Winding,Ying-JC-winding,Ying-Topo-JC-nonHermitian}
have recently arised as a special focus. Applications for critical
quantum metrology~\cite%
{Garbe2020,Garbe2021-Metrology,Ilias2022-Metrology,Ying2022-Metrology,Hotter2024-Metrology} have
been proposed, with a great potential to become practical
ultra-high-precision quantum technology in the contemporary era of
ultra-strong coupling
~\cite{Ciuti2005EarlyUSC,Aji2009EarlyUSC,Diaz2019RevModPhy,Kockum2019NRP,Wallraff2004,Gunter2009,
Niemczyk2010,Peropadre2010,FornDiaz2017,Forn-Diaz2010,Scalari2012,Xiang2013,Yoshihara2017NatPhys,Kockum2017,
Ulstrong-JC-2,Ulstrong-JC-3-Adam-2019}
and even deep-strong coupling~\cite{Yoshihara2017NatPhys,Bayer2017DeepStrong,Ulstrong-JC-1}.

A most fundamental prototype of light-matter interaction is the quantum Rabi model (QRM)~\cite{rabi1936,Rabi-Braak,Eckle-Book-Models} which contains both the rotating-wave terms~\cite{JC-model,JC-Larson2021} and counter-rotating terms. In the ultra-strong coupling regime, it has been experimentally found that the counter-rotating terms are playing an indispensable role\cite{PRX-Xie-Anistropy,Forn-Diaz2010},
which brings more attention to the QRM. Theoretically, the milestone work~\cite{Braak2011} revealing the
integrability of the QRM has inspired a massive dialogue~\cite{Solano2011}
between mathematics and physics in light-matter interactions~\cite{
Braak2011,Solano2011,Boite2020,Liu2021AQT,Diaz2019RevModPhy,Kockum2019NRP,Rabi-Braak,Braak2019Symmetry,
Wolf2012,FelicettiPRL2020,Felicetti2018-mixed-TPP-SPP,Felicetti2015-TwoPhotonProcess,Simone2018,Alushi2023PRX,
Irish2014,Irish2017,Irish-class-quan-corresp,JC-Larson2021,
PRX-Xie-Anistropy,Batchelor2015,XieQ-2017JPA,
Hwang2015PRL,Bera2014Polaron,Hwang2016PRL,Ying2015,LiuM2017PRL,Ying-2018-arxiv,Ying-2021-AQT,Ying-gapped-top,Ying-Stark-top,Ying-Spin-Winding,
Ying-JC-winding,Ying-Topo-JC-nonHermitian,Grimaudo2022q2QPT,Grimaudo2023-Entropy,
CongLei2017,CongLei2019,Ying2020-nonlinear-bias,LiuGang2023,ChenQH2012,Zhu-PRL-2020,
e-collpase-Garbe-2017,e-collpase-Duan-2016,Garbe2020,Rico2020,
Garbe2021-Metrology,Ilias2022-Metrology,Ying2022-Metrology,
Boite2016-Photon-Blockade,Ridolfo2012-Photon-Blockade,Li2020conical,
Ma2020Nonlinear,
Padilla2022,ZhengHang2017,Yan2023-AQT,Zheng2017,Chen-2021-NC,Lu-2018-1,Gao2021,PengJie2019,Liu2015,Ashhab2013,Yimin2018, ChenGang2012,FengMang2013,Eckle-2017JPA,Maciejewski-Stark,Xie2019-Stark,Casanova2018npj,HiddenSymMangazeev2021,HiddenSymLi2021,HiddenSymBustos2021,
JC-Larson2021,Stark-Cong2020,Cong2022Peter,Stark-Grimsmo2013,Stark-Grimsmo2014,Downing2014,Ulstrong-JC-2,Chen2018,WangZJ2020PRL}. A fascinating
feature of the QRM is that as a few-body system it possesses~\cite{Liu2021AQT,Ashhab2013,Ying2015,Hwang2015PRL,Ying2020-nonlinear-bias,Ying-2021-AQT,LiuM2017PRL,Hwang2016PRL,Irish2017,
Ying-gapped-top,Ying-Stark-top,Ying-2018-arxiv,Ying-Spin-Winding,Grimaudo2022q2QPT} a quantum phase transition (QPT)\cite{Sachdev-QPT} which traditionally
lies in condensed matter, despite that it might be a matter of taste to term
the transition quantum or not by considering the negligible quantum
fluctuations in the photon vacuum state~\cite{Irish2017}.
Indeed, the QPT in the QRM can be bridged to the thermodynamical limit via the universality of the critical exponent~\cite{LiuM2017PRL}. Moreover, extended versions of the QRM with anisotropy~\cite{PRX-Xie-Anistropy,LiuM2017PRL,
Ying-2021-AQT,Ying-Stark-top,Ying-gapped-top,Ying-Spin-Winding,
Yimin2018,Pietikainen2017,Wang2019Anisotropy} and Stark nonlinear coupling~\cite{Eckle-2017JPA,Maciejewski-Stark,Ying-Stark-top,Xie2019-Stark,Ying-JC-winding}
also manifest single-qubit topological phase transitions~\cite{Ying-2021-AQT,Ying-gapped-top,Ying-Stark-top,Ying-Spin-Winding}, including both conventional ones~\cite{Ying-2021-AQT,Ying-gapped-top,Ying-Stark-top,Ying-Spin-Winding} and unconventional ones~\cite{Ying-gapped-top,Ying-Stark-top,Ying-Spin-Winding}, analogously to those in condensed matter~\cite{Topo-KT-transition,Topo-KT-NoSymBreak,Topo-Haldane-1,Topo-Haldane-2,Topo-Wen,ColloqTopoWen2010,Amaricci-2015-no-gap-closing,Xie-QAH-2021,
Topo-Wen,Hasan2010-RMP-topo,Yu2010ScienceHall,Chen2019GapClosing,TopCriterion,Top-Guan,TopNori}. The conventional topological phase transitions occur with gap closing, while the unconventional ones emerge unexpectedly in gapped situations either in level anti-crossing~\cite{Ying-Spin-Winding,Ying-JC-winding} or completely without any tendency of gap closing~\cite{Ying-gapped-top,Ying-Stark-top}. At this point it may be worthwhile to mention that finite-size QPTs can also occur with level crossing in other fields, e.g. in pairing-depairing models~\cite{Ying2008PRL,Ying2007confined,Ying2006SCFM} and coupled fermion-boson models~\cite{Stojanovic2020PRB,Stojanovic2020PRL,Stojanovic2021PRA} where one also encounters transitions of topological structure in the energy spectrum~\cite{Ying2008PRL,Ying2007confined,Ying2006SCFM} and the real space~\cite{Ying2008PRL}.
As a matter of fact, the finite system in lighter-matter interaction forms a mini-world of phase transitions~\cite{Ying-gapped-top}. The abundant physics in such a mini-world also brings conceptional renovations, e.g., universality and
diversity as antagonists by nature can counter-intuitively support each other~\cite{Ying-gapped-top,Ying-Stark-top}; it is also found that the conventionally incompatible symmetry-breaking Landau class of transitions\cite{Landau1937} and symmetry-protected
topological class of transitions~\cite{Topo-KT-transition,Topo-KT-NoSymBreak,Topo-Haldane-1,Topo-Haldane-2,Topo-Wen,ColloqTopoWen2010} can coexist in a same system and even occur simultaneously~\cite{Ying-Stark-top,Ying-JC-winding}. Favorably for practical applications the topological features are robust against the non-Hermiticity induced by decay rates and dissipation~\cite{Ying-Topo-JC-nonHermitian}.

Unlike the topological phase transitions which emerge at finite frequencies~\cite{Ying-2021-AQT,Ying-gapped-top,Ying-Stark-top,Ying-Spin-Winding}, the QPT of the QRM occurs at low frequencies~\cite{Liu2021AQT,Ashhab2013,Ying2015,Hwang2015PRL,Ying2020-nonlinear-bias,Ying-2021-AQT,LiuM2017PRL,Irish2017,
Ying-gapped-top,Ying-Stark-top,Ying-2018-arxiv,Grimaudo2022q2QPT}. The conventional expression for the QPT of the QRM was obtained in semiclassical approximation~\cite{Ashhab2013,Ying-2018-arxiv,Ying2020-nonlinear-bias} which is valid only in the extreme limit of low frequency. However, in reality one always has a finite frequency in the practical systems.
It has been found that the transition will shift away from the conventional transition coupling once one goes away from zero frequency and an improved scale of transition coupling was proposed~\cite{Ying2015}. Although the new scale of transition coupling can qualitatively better capture the transition coupling, a quantitatively accurate expression for the transition coupling is still lacking. On the other hand, the QPT of the QRM has been applied for critical quantum metrology~\cite{Garbe2020,Garbe2021-Metrology,Ilias2022-Metrology,Ying2022-Metrology} to achieve
high precision bound of experimental measurements represented by the quantum Fisher information (QFI)~\cite{Cramer-Rao-bound,Taddei2013FisherInfo,RamsPRX2018,Zhou-FidelityQPT-2008,Gu-FidelityQPT-2010}. Actually such applications depend on the knowledge of the exact location of the transition coupling beforehand, which in turn raises the requirement to know the accurate location of transition coupling. In such a situation, an accurate expression for the frequency dependence of the transition coupling of the QRM is highly desirable.

In the present work we conversely use the QFI in the quantum metrology to identify the transition coupling of the QRM.  The result from the QFI shows that the transition couplings of the QRM indeed much deviate from the conventional expression of the transition coupling once a finite frequency is turned on. Besides the numerics by the exact diagonalization on the QFI, we also formulate the QFI in the variational polaron picture\cite{Ying2015} to analytically reproduce the numeric results of the QFI. An accurate expression for the transition coupling is obtained by the inspiration from the fractional-power-law effect of polaron frequency renormalization. From the QFI in the polaron picture we find that the transition occurs around a point where the effective polaron velocity and the susceptibility of the single-photon absorption rate reach their maxima. Our results provide an accurate reference of transition couplings for quantum metrology at non-zero frequencies. The QFI in the polaron picture also provides an accurate compensation for the parameter regime where the exact diagonalization can not access due to demanding requirement of large basis cutoff. Our analysis may also add some insight for the bridge of the QFI and the transition properties.

The paper is organized as follows. Section \ref{Sect-Model} introduces the
QRM. Section \ref{Sect-gC0-gC1} shows the deviation of the conventional transition coupling at non-zero frequencies. Section \ref{Sect-QFI} presents the QFI with numerical identification and analytical expressions for accurate transition coupling. Section \ref{Sect-QFI-in-PP} is devoted for formulation of the QFI in the polaron picture. The general expression of the QFI is simplified by a finding of a vanishing term. Maximum effective velocity is shown around the transition identified by the QFI. Section \ref{Sect-Conclusions} demonstrates the transition coincidence with the maximum susceptibility of single-photon absorption rate.
Finally Section \ref{Sect-Conclusions} gives a summary of conclusions.

\section{Model and symmetry}

\label{Sect-Model}

The QRM \cite{rabi1936,Rabi-Braak,Eckle-Book-Models} takes the following form%
\begin{equation}
H=\omega a^{\dagger }a+g\sigma _{z}(a^{\dagger }+a)+\frac{\Omega }{2}\sigma
_{x}
\end{equation}%
which describes the coupling between a bosonic mode with frequency $\omega $
and a qubit represented by the Pauli matrices $\sigma _{x,y,z}.$ Here we
have adopted the spin notation as in Ref.\cite{Irish2014}, in which $\sigma
_{z}=\pm $ conveniently represents the two flux states in the flux-qubit
circuit system\cite{flux-qubit-Mooij-1999}, while superconducting circuit
systems can realize ultra-strong coupling~\cite{Ciuti2005EarlyUSC,Aji2009EarlyUSC,Diaz2019RevModPhy,Kockum2019NRP,Wallraff2004,Gunter2009,Niemczyk2010,Peropadre2010,FornDiaz2017,Forn-Diaz2010,Scalari2012,Xiang2013,Yoshihara2017NatPhys,Kockum2017}
 and even deep-strong coupling~\cite{Yoshihara2017NatPhys,Bayer2017DeepStrong},
respectively with coupling strength $g$ beyond $0.1\omega $ and $1.0\omega $. The
operator $a^{\dagger }$ ($a$) creates (annihilates) a boson and $a^{\dagger
}a$ is the boson (photon) number. In the conventional spin notation $\Omega $
is the level splitting. One can retrieve the conventional notation by a spin
rotation \{$\sigma _{x},\sigma _{y},\sigma _{z}$\} $\rightarrow $ \{$\sigma
_{z},-\sigma _{y},\sigma _{x}$\} around the axis $\vec{x}+\vec{z}$.

By transformation $a^{\dagger }=(\hat{x}-i\hat{p})/\sqrt{2},$ $a=(\hat{x}+i%
\hat{p})/\sqrt{2}$, where $x$ and $\hat{p}=-i\frac{\partial }{\partial x}$ are the effective position and momentum, we can
rewrite the Hamiltonian in position space%
\begin{equation}
H=\sum _{\sigma _{z}=\pm }h_{\sigma _{z}}\left\vert \sigma _{z}\right\rangle
\left\langle \sigma _{z}\right\vert +\frac{\Omega }{2}
\sum _{\sigma _{z}=\pm }
\left\vert \sigma _{z}\right\rangle \left\langle
\overline{\sigma }_{z}\right\vert  \label{Hx}
\end{equation}%
where $\sigma _{z}=-\overline{\sigma }_{z}=\pm $ represents the spin state
in $z$ direction. Here $h_{\sigma _{z}}=\frac{\omega }{2}\hat{p}%
^{2}+v_{\sigma _{z}}\left( x\right) $ is effective singe-particle
Hamiltonian in the spin-dependent potential $v_{\sigma _{z}}(x)=\omega
\left( x+\widetilde{g}\sigma _{z}\right) ^{2}/2+\varepsilon _{0}^{z}$ with a
constant energy $\varepsilon _{0}^{z}=-\frac{1}{2}[\widetilde{g}%
^{2}+1]\omega $. Indeed $x$ and $p$ can be represented by
the flux and charge of Josephson junctions in circuit systems~\cite{flux-qubit-Mooij-1999,you024532}. In such a representation the coupling plays a role to
separate the potential in opposite directions with a displacement denoted
by the rescaled coupling $\widetilde{g}=\sqrt{2}g/\omega $. The $\Omega $
term now acts as spin flipping in $\sigma _{z}$ space or tunneling in
position space \cite{Ying2015,Irish2014}.

The model possesses the parity
symmetry $\hat{P}=\sigma _{x}(-1)^{a^{\dagger }a}$ which commutes with
the Hamiltonian. The parity symmetry leads to an antisymmetric ground-state wave function under exchange of spin and inversion of position simultaneously, which will simplify our formulation.

\section{Conventional transition coupling and deviations at non-zero
frequencies}

\label{Sect-gC0-gC1}

The QRM manifests a quantum phase transition\cite%
{Liu2021AQT,Ashhab2013,Ying2015,Hwang2015PRL,Ying2020-nonlinear-bias,Ying-2021-AQT,LiuM2017PRL,Hwang2016PRL,Irish2017,Ying-gapped-top,Ying-Stark-top,Ying-2018-arxiv,Ying-Spin-Winding,Grimaudo2022q2QPT}
at a critical point which has a conventional location at
\begin{equation}
g_{c0}=\frac{\sqrt{\omega \Omega }}{2}.  \label{gc0}
\end{equation}%
Note here that $g_{c0}$ is frequency-dependent. However, in reality this
expression of $g_{c0}$ is exact only in the extreme limit of low frequency.
It has been found that the transition will shift away from $g_{c0}$ at
finite frequencies and a new scale of transition coupling was proposed as~\cite{Ying2015}
\begin{equation}
g_{c1}=\sqrt{\omega ^{2}+\sqrt{\omega ^{4}+g_{c0}^{4}}}  \label{gc1}
\end{equation}%
which better captures the transition at finite frequencies.

Figures \ref{fig-gc}(a) and \ref{fig-gc}(b) compare the frequency dependence
of $g_{c0}$ (orange dotted line) and $g_{c1}$ (green (light gray) solid
line) in linear scale and logarithmic scale of the frequency ratio $\omega
/\Omega $. From Fig. \ref{fig-gc}(b) one sees that $g_{c0}$ and $g_{c1}$
coincide at $\omega /\Omega =0$, while they depart from each other
immediately once the frequency is turned on to any finite value. This
indicates that $g_{c0}$ may be inaccurate even at the very low frequencies
except right at zero frequency, which turns out to be true as in comparison
with the result of the QFI (dots, later addressed in next section). In practice,
the bosonic mode in the light-matter interaction has a non-zero frequency,
the inaccuracy of $g_{c0}$ at non-zero frequencies would hinder the
application, especially when high precision is the goal of quantum metrology which in turn requires the knowledge
of accurate location of the transition coupling.
On the other hand, although $g_{c1}$ yields a qualitative
improvement in scale estimation of the transition coupling, a quantitatively more accurate
transition coupling is still lacking. We shall obtain an accurate transition
coupling $g_{c2}$ by a combination of the QFI and the polaron picture in the
following section.

\section{Quantum Fisher information (QFI) and accurate frequency dependence of transition coupling}
\label{Sect-QFI}

\subsection{QFI for quantum phase transition}

In measurements the precision of any experimental estimation of the
parameter $\lambda $ in the Hamiltonian is bounded by $F_{Q}^{1/2}$\cite%
{Cramer-Rao-bound}. Here $F_{Q}$ is the QFI \cite%
{Cramer-Rao-bound,Taddei2013FisherInfo,RamsPRX2018} which takes the
following form for pure states
\begin{equation}
F_{Q}=4\left[ \langle \psi ^{\prime }\left( \lambda \right) |\psi ^{\prime
}\left( \lambda \right) \rangle -\left\vert \langle \psi ^{\prime }\left(
\lambda \right) |\psi \left( \lambda \right) \rangle \right\vert ^{2}\right]
,  \label{QFI}
\end{equation}%
where $^{\prime }$ denotes the derivative of the ground state $|\psi
(\lambda )\rangle $ of $H$ with respect to $\lambda $. So the QFI is the
precision criteria quantity in quantum metrology, with higher QFI meaning
higher measurement precision. Around a QPT the ground-state wave function is
varying quickly, which provides a sensitivity resource for the so-called
critical quantum metrology\cite%
{Garbe2020,Garbe2021-Metrology,Ilias2022-Metrology,Ying2022-Metrology}, with
the a maximum QFI (thus a maximum measurement precision) available at the
QPT.

On the other hand, actually $\chi _{F}=F_{Q}/4$ is the susceptibility of the
fidelity~\cite{Gu-FidelityQPT-2010,You-FidelityQPT-2007,You-FidelityQPT-2015}
\begin{equation}
F=\left\vert \langle \psi \left( \lambda \right) |\psi \left( \lambda
+\delta \lambda \right) \rangle \right\vert =1-\frac{\delta \lambda ^{2}}{2}%
\chi _{F}
\end{equation}%
in an infinitesimal variation $\delta \lambda $ of the parameter $\lambda $.
The fidelity can be a basic quantity to characterize the QPT \cite{Zhou-FidelityQPT-2008,Zanardi-FidelityQPT-2006,Gu-FidelityQPT-2010,You-FidelityQPT-2007,You-FidelityQPT-2015}, with a peak of the fidelity susceptibility being a transition
signal.

\subsection{QFI of the QRM}

\begin{figure*}[t]
\includegraphics[width=2\columnwidth]{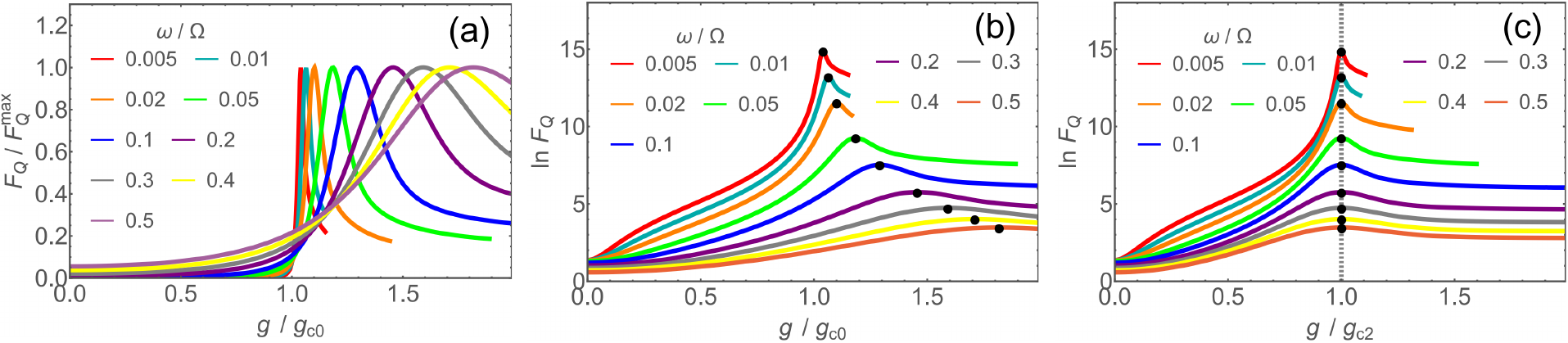}
\caption{Quantum Fisher information $F_Q$ with respect to coupling $g$ at different frequency ratios $\omega/\Omega$.
(a) $F_Q$ renormalized by its maximum (peak value) $F_Q^{\rm max}$.
(b) $F_Q$ with maximum $F_Q^{\rm max}$ (black dots) in natural logarithm.
(c) $F_Q$ with maximum positions (black dots) re-aligned in a vertical line in a scaling $g/g_{c2}$ by the accurate transition coupling $g_{c2}$ in \eqref{gc2}.
Here $F_Q$ is numerically obtained by exact diagonalization (ED) for $\omega/\Omega=0.005,0.01,0.02,0.05,0.1,0.2,0.3,0.4,0.5$ from left to right in (a) and from upper to lower in (b).
}
\label{fig-Fq-w}
\end{figure*}

From both points of view of the critical quantum metrology and the
fidelity theory of the QPT, we can utilize the QFI $F_{Q}$ to identify the QPT
in the QRM. Here the coupling $g$ is taken as the parameter for $\lambda $.
Figure \ref{fig-Fq-w} shows the evolution of $F_{Q}$, numerically
obtained by the method of exact diagonallization (ED) \cite%
{Ying2020-nonlinear-bias,Ying-Spin-Winding}, in the variation of the
coupling $g$. Here in panel (a) of the figure $F_{Q}$ is scaled by its
maximum $F_{Q}^{{\rm \max }}$ at each fixed frequency ratio $\omega /\Omega $%
. We see that $F_{Q}$ indeed has a peak, which is located at $g_{c0}$ in the
extreme limit of low frequency $\omega /\Omega \rightarrow 0$ but moves to
larger relative coupling $\overline{g}=g/g_{c0}$ once the frequency is
raised from zero. We denote as $g_{c{\rm F}}$ for the transition coupling
identified by the location of $F_{Q}^{{\rm \max }}.$ The result confirms the
inaccuracy of $g_{c0}$ and the shift of the transition away from $g_{c0}$
for any finite frequency. The evolutions of the peak position and the peak
value $F_{Q}^{{\rm \max }}$ can be seen more clearly from the black dots in
Fig. \ref{fig-Fq-w}b where $F_{Q}$ is un-rescaled. Here the logarithmic plot
of $F_{Q}$ indicates the large values of the QFI around $F_{Q}^{{\rm \max }}$%
, which is favorable for the quantum metrology.

\begin{figure*}[t]
\includegraphics[width=2\columnwidth]{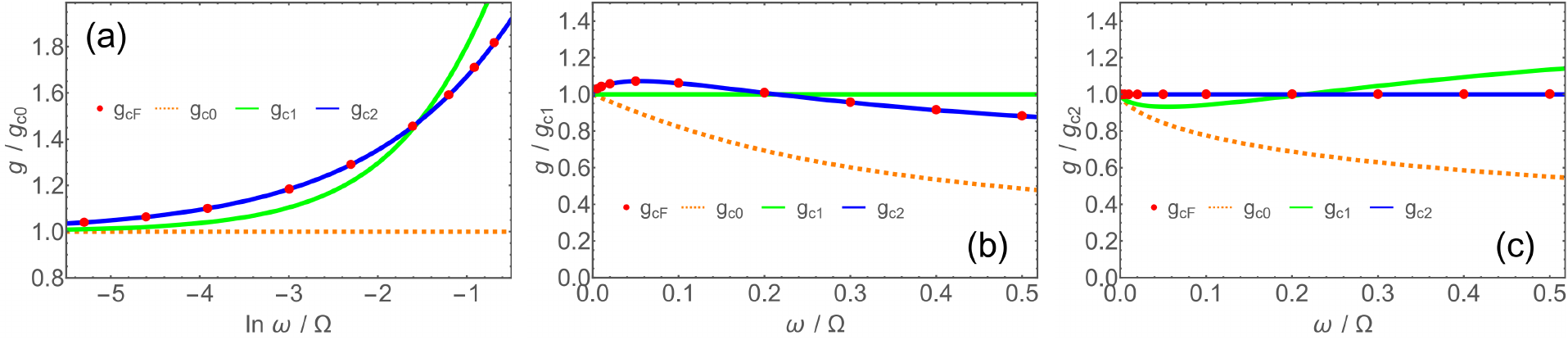}
\caption{Frequency dependence of transition coupling $g_c$ in different expressions. (a) $g_c$ scaled by $g_{c0}$ in natural logarithm of frequency ratio $\omega/\Omega$. (b) $g_c$ scaled by $g_{c1}$ versus $\omega/\Omega$. (c) $g_c$ scaled by $g_{c2}$. Here $g_{\rm cF}$ (black dots) as the reference of accurate transition coupling is numerically extracted from position of $F_Q^{\rm max}$, while $g_{c0}$ (orange dotted line) is the conventional transition coupling in \eqref{gc0}, $g_{c1}$ (green (light gray) solid line) is the transition coupling in polaron picture in \eqref{gc1}, and finally $g_{c2}$ (blue (dark gray) solid line) is the analytical transition coupling obtained from $F_Q^{\rm max}$ in \eqref{gc2}.
}
\label{fig-gc}
\end{figure*}

\subsection{Accurate transition coupling of the QRM}
\label{Sect-gc2-accurate}

As afore-mentioned in Sec.\ref{Sect-gC0-gC1}, the frequency dependence of
the conventional transition coupling $g_{c0}$ has large deviations from the
beginning in going away from zero frequency and the improved one $g_{c1}$ is
not either accurate enough even at low frequencies. Actually $g_{c0}$ can be
equivalently obtained by a semiclassical approximation which assumes a mass
of point as the classical particle in the effective field for the spatial
part while keeping quantum character for the spin part\cite%
{Ying2020-nonlinear-bias,Ying-2018-arxiv}. It is the classical treatment in
the spatial part that leads to the deviation, since in full quantum
mechanical picture the particle should appear as a wave packet for spatial
distribution instead of a point of mass. As an improvement $g_{c1}$
considers the spatial wave packet in the polaron picture, however as an
analytical simplification the frequency renormalization of the polarons has
not been included~\cite{Ying2015}.

Indeed, as demonstrated in Appendix \ref{Appendix-gC-ksi}, when the
frequency renormalization of the polarons is picked up one gets a correct
direction for further improvement of the transition coupling as $g_{c\xi }$.
It turns out that the frequency renormalization effect enters in the
improvement of the transition coupling in a form of fractional power low $%
\omega ^{2n/3}$ of the frequency as in (\ref{gC-kesi-fract-power-law}).

Although $g_{c\xi }$ is still not accurate enough due to the analytical
simplification, it inspires us to propose accurate expressions with adjusted
coefficients from $g_{c\xi }$. To the second order we find
\begin{equation}
g_{c2}=g_{c0}\left[ 1+\frac{1}{100\alpha _{{\rm FS}}}(\frac{\omega }{\Omega }%
)^{2/3}-\frac{1}{8}(\frac{\omega }{\Omega })^{4/3}\right] ,  \label{gc2}
\end{equation}%
where $\alpha _{{\rm FS}}=1/137$ by coincidence is the fundamental
fine-structure constant in quantum mechanics, works very accurately in $%
\omega /\Omega \in \lbrack 0,0.5]$ regime by comparison with the numerical $%
g_{c{\rm F}}$, as one sees in Fig. \ref{fig-gc}(a-c) and Fig. \ref{fig-Fq-w}%
(c). Another second-order expression
\begin{equation}
g_{c2}=g_{c0}\left[ 1+\frac{4}{3}(\frac{\omega }{\Omega })^{2/3}-\frac{3}{40}%
(\frac{\omega }{\Omega })^{4/3}\right] ,  \label{gc2-34}
\end{equation}%
can extend the validity up to $\omega /\Omega =3.0$ with a small price of
larger error than (\ref{gc2}) in order $10^{-3}$ in $\omega /\Omega \in
\lbrack 0,0.5]$. These expressions are extracted by coefficient
fractionization in second-order least-squares fitting. If one keeps
decimalized coefficients, a fitting expression in the third order
\begin{equation}
g_{c2}^{{\rm fitting}}=g_{c0}\left[ 1+c_{1}(\frac{\omega }{\Omega }%
)^{2/3}+c_{2}(\frac{\omega }{\Omega })^{4/3}+c_{3}(\frac{\omega }{\Omega }%
)^{6/3}\right] ,  \label{gc-fitting}
\end{equation}%
with $c_{1}=1.3715$, $c_{2}=-0.1311$, $c_{3}=0.0184$ can capture the maximum
point of $F_{Q}$ accurately in the entire frequency regime up to $\omega
/\Omega =3.0$, though the transition becomes a crossover in high frequencies.
A comparison with the ED result in low-frequency regime and high-frequency
regime is provided in Appendix \ref{Appendix-compare-gc2}. Despite that $%
g_{c2}^{{\rm fitting}}$ has a wider range for accuracy of the position of $%
F_{Q}^{{\rm \max }}$, the expression $g_{c2}$ in (\ref{gc2}) and (\ref%
{gc2-34}) is easier to remember with the fractionized coefficients (rather
than decimalized) and it has the same accuracy as $g_{c2}^{{\rm fitting}}$
in the low-frequency regime where the transition makes more sense.

At this point one might speculate that a Fourier expansion with integer
powers of the frequency $(\omega/\Omega)^n$ should also work without need of
the fractional-power form. However, as demonstrated in Appendix \ref%
{Appendix-Fourier-Expansion}, such a Fourier-expansion fitting is very
inefficient while fractional-power fitting already reaches a good
convergence with only a couple of orders, without mentioning that such a
Fourier expansion does not provide any physical insight if it were not
coinciding with integer-power-series law of $g_{c1}$ as in Eq.~%
\eqref{gc1-Expansion}. This indicates that the integer-power law in the
Fourier expansion does not capture the physical essence of frequency
renormalization around the transition in the QRM.

The accurate transition coupling $g_{c2}$ should be useful in applications
of the quantum metrology as in practice the frequency is non-zero. Even if
the transition becomes more broadened at a higher frequency, $g_{c2}$ still
can provide an accurate location for the maximum QFI where one can find
relatively the highest measurement precision in variation of the coupling
despite that one does not achieve an ideal divergence.

\section{Polaron picture for QFI}
\label{Sect-QFI-in-PP}

As afore-mentioned, the above result of the QFI\ at finite frequencies is
numerically obtained by the method of ED \cite%
{Ying2020-nonlinear-bias,Ying-Spin-Winding} which needs a cutoff of the
basis number. In the presence of a large coupling at low frequencies such a
basis number might be divergently large, which might lead to a heavy
computation cost and even goes beyond the numerical access limit. In such a
situation an analytical form for the QFI would be desirable. The variational
method of polaron picture (PP) \cite{Ying2015} provides such a possibility
with high accuracy over the entire coupling regime including the transition
area. The validity of the PP is not limited by the frequency and it often
facilitates the understanding of underlying physics in light-matter
interactions \cite%
{Ying2015,CongLei2017,CongLei2019,Ying2020-nonlinear-bias,Ying-2021-AQT,Ying-gapped-top,Ying-Stark-top,Liu2021AQT,Ying-2018-arxiv}%
. In this section we shall formulate the QFI of the QRM in the PP.

\subsection{Formulation of QFI in polaron picture}

The eigen wave function of the QRM has two spin components
\begin{equation}
\left\vert \psi \left( \overline{g}\right) \right\rangle =\frac{1}{\sqrt{2}}%
\sum_{\sigma _{z}=\pm }\psi _{\sigma _{z}}\left( \overline{g}\right)
\left\vert \sigma _{z}\right\rangle
\end{equation}%
subject to the normalization condition
\begin{equation}
\left\langle \psi \left( \overline{g}\right) |\psi \left( \overline{g}%
\right) \right\rangle =\frac{1}{2}\sum_{\sigma _{z}=\pm }\left\langle \psi
_{\sigma _{z}}\left( \overline{g}\right) |\psi _{\sigma _{z}}\left(
\overline{g}\right) \right\rangle =1.  \label{normalization-condition}
\end{equation}%
Equivalently $\left\langle \psi _{\sigma _{z}}\left( \overline{g}\right)
|\psi _{\sigma _{z}}\left( \overline{g}\right) \right\rangle =1$ and the
inner product can be written in integral form $\left\langle \psi  _{\sigma
_{z}}\left( \overline{g}\right) |\psi _{\sigma _{z}}\left( \overline{g}%
\right) \right\rangle =\int dx\psi ^* _{\sigma _{z}}\left( \overline{g}%
,x\right) \psi _{\sigma _{z}}\left( \overline{g},x\right) $ in the position
space. Here we have defined the scaled coupling $\overline{g}=g/g_{c0}$,
while the QFI with respect to $g$ simply differs by a factor
\begin{equation}
F_{Q}\left( g\right) =F_{Q}\left( \overline{g}\right) \left( \frac{d%
\overline{g}}{dg}\right) ^{2}=\frac{1}{g_{c0}^{2}}F_{Q}\left( \overline{g}%
\right) .
\end{equation}%
In the following we formulate the QFI $F_{Q}\left( \overline{g}\right) $
with respect to $\overline{g}$ in the PP.

In the variational PP each spin component of the wave function can be
decomposed into a linear combination of polarons represented by $\varphi _{i}
$
\begin{eqnarray}
\psi _{+}\left( \overline{g},x\right)  &=&\sum_{i=\alpha ,\beta }w_{i}\left(
\overline{g}\right) \varphi _{i}\left( \overline{g},x\right)  \\
\psi _{-}\left( \overline{g},x\right)  &=&P\psi _{+}\left( \overline{g}%
,-x\right)
\end{eqnarray}%
where $P=-1$ is the negative parity for the ground state. Explicitly the
polaron in the ground state takes the form of Gaussian wave packet
\begin{equation}
\varphi _{i}\left( \overline{g},x\right) =\left( \xi _{i}/\pi \right)
^{1/4}\exp \left[ -\xi _{i}\left( x+x_{i}\right) ^{2}/2\right]
\label{polaron-Wave-Function}
\end{equation}%
with polaron displacement $x_{i}=\zeta _{i}g^{\prime }$, as renormalized
from the potential displacement $g^{\prime }$ by $\zeta _{i},$ and the
frequency renormalization factor $\xi _i $~\cite{Ying2015}. The frequency
renormalization gives a more compact representation of polarons than the
coherent state expansion\cite{Bera2014Polaron} and in Appendix \ref%
{Appendix-gC-ksi} we also see that the frequency renormalization yields an
correct direction for improvement of the transition coupling.

Here we have
adopted the two-polaron decomposition \cite{Ying2015} for the ground state
of the QRM\ with the weights%
\begin{equation}
w_{\alpha }\left( \overline{g}\right) =\alpha ,\quad w_{\beta }\left(
\overline{g}\right) =\beta .
\end{equation}%
The two-polaron decomposition with frequency renormalization is already
accurate enough for our discussion and convenient for physical descriptions,
although extension to multi-polaron representation is direct and might gives
some small quantitative improvements\cite{CongLei2017}.

The two terms in $%
F_{Q}$ are then obtained by
\begin{eqnarray}
&&\langle \psi ^{\prime }\left( \overline{g}\right) |\psi \left( \overline{g}%
\right) \rangle   \nonumber \\
&=&\sum_{i,j}\left[ w_{i}\left( \overline{g}\right) w_{j}\left( \overline{g}%
\right) \langle \varphi _{i}^{\prime }|\varphi _{j}\rangle +w_{i}^{\prime
}\left( \overline{g}\right) w_{j}\left( \overline{g}\right) \langle \varphi
_{i}|\varphi _{j}\rangle \right] ,
\end{eqnarray}%
and%
\begin{eqnarray}
&&\langle \psi ^{\prime }\left( \overline{g}\right) |\psi ^{\prime }\left(
\overline{g}\right) \rangle   \nonumber \\
&=&\sum_{i,j}\left[ w_{i}\left( \overline{g}\right) w_{j}\left( \overline{g}%
\right) \langle \varphi _{i}^{\prime }|\varphi _{j}^{\prime }\rangle
+w_{i}^{\prime }\left( \overline{g}\right) w_{j}^{\prime }\left( \overline{g}%
\right) \langle \varphi _{i}|\varphi _{j}\rangle \right]   \nonumber \\
&&+\sum_{i,j}\left[ w_{i}\left( \overline{g}\right) w_{j}^{\prime }\left(
\overline{g}\right) \langle \varphi _{i}^{\prime }|\varphi _{j}\rangle
+w_{i}^{\prime }\left( \overline{g}\right) w_{j}\left( \lambda \right)
\langle \varphi _{i}|\varphi _{j}^{\prime }\rangle \right] .
\end{eqnarray}%
The first-order derivative terms of the inter products include the
variations of displacemnent and frequency renormalization with respect to
the coupling%
\begin{equation}
\langle \varphi _{i}^{\prime }|\varphi _{j}\rangle =\langle \frac{\partial
\varphi _{i}}{\partial x_{i}}|\varphi _{j}\rangle \frac{dx_{i}}{d\overline{g}%
}+\langle \frac{\partial \varphi _{i}}{\partial \xi _{i}}|\varphi
_{j}\rangle \frac{d\xi _{i}}{d\overline{g}}. \label{1st-derivative-terms}
\end{equation}%
and the second-order derivative terms collect their quadratic mixture%
\begin{eqnarray}
\langle \varphi _{i}^{\prime }|\varphi _{j}^{\prime }\rangle  &=&\langle
\frac{\partial \varphi _{i}}{\partial x_{i}}|\frac{\partial \varphi _{j}}{%
\partial x_{j}}\rangle \frac{dx_{i}}{d\overline{g}}\frac{dx_{j}}{d\overline{g%
}}+\langle \frac{\partial \varphi _{i}}{\partial x_{i}}|\frac{\partial
\varphi _{j}}{\partial \xi _{j}}\rangle \frac{dx_{i}}{d\overline{g}}\frac{%
d\xi _{j}}{d\overline{g}}  \nonumber \\
&&+\langle \frac{\partial \varphi _{i}}{\partial \xi _{i}}|\frac{\partial
\varphi _{j}}{\partial x_{j}}\rangle \frac{d\xi _{i}}{d\overline{g}}\frac{%
dx_{j}}{d\overline{g}}+\langle \frac{\partial \varphi _{i}}{\partial \xi _{i}%
}|\frac{\partial \varphi _{j}}{\partial \xi _{j}}\rangle \frac{d\xi _{i}}{d%
\overline{g}}\frac{d\xi _{j}}{d\overline{g}}. \label{2nd-derivative-terms}
\end{eqnarray}%
The expressions for $\langle \varphi _{i}^{\prime }|\varphi _{j}\rangle ,$ $%
\langle \varphi _{i}^{\prime }|\varphi _{j}^{\prime }\rangle $, $dx_{i}/d%
\overline{g}$ and $d\xi _{i}/d\overline{g}$ are explicitly available as
provided in Appendix \ref{Appendix-Fq-PP-expressions}.

\begin{figure}[t]
\includegraphics[width=1\columnwidth]{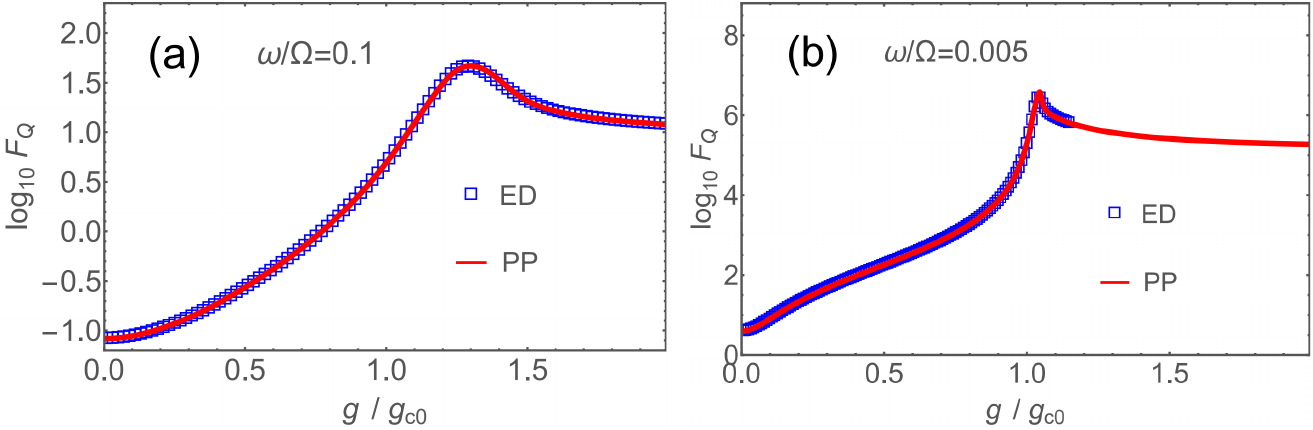}
\caption{Comparison of $F_Q$ for the ED (symbols) and the variational method of polaron picture (PP) (solid lines). (a) $\omega/\Omega=0.1$. (b) $\omega/\Omega=0.005$. The missing data in (b) indicates the numerical difficulty for the ED in low frequencies and large couplings, while it is readily accessible and compensated by the variational PP.
}
\label{fig-Compare-ED-PP}
\end{figure}

\subsection{Accuracy of QFI in polaron picture and compensation for ED}

The result of the QFI in the PP is quite accurate and also provides a
compensation for the ED in case that the ED leads to an overload of computation cost due to the need
of large cutoff of basis number. Indeed, as illustrated by Fig.\ref%
{fig-Compare-ED-PP}, the PP (solid lines) basically reproduces the ED
results (circles) of the QFI in the entire regime of coupling. Note here the
ED can go deep into the large coupling regime beyond the transition for the
frequency $\omega /\Omega =0.1$ in Fig.\ref{fig-Compare-ED-PP}(a), while
access to large couplings by the ED might become difficult at low
frequencies as indicated by the missing data at $\omega /\Omega =0.05$ in
Fig.\ref{fig-Compare-ED-PP}(b). Nevertheless, with the accuracy guaranteed,
as checked in the ED-accessible regime, the missing data in the inaccessible
regime for the ED can be readily compensated by the result of the PP as
demonstrated by the extended solid line in Fig.\ref{fig-Compare-ED-PP}(b).
Thus, the formulation of the QFI in the PP provides an accurate and
convenient tool without heavy cost of computation to cover the entire
parameter regime.

\subsection{Vanishing $\langle \protect\psi ^{\prime }\left( \protect\lambda %
\right) |\protect\psi \left( \protect\lambda \right) \rangle $ term}

From the convenient analysis in the PP we come across the vanishing of the
first derivative term in the QFI
\begin{equation}
\langle \psi ^{\prime }\left( \overline{g}\right) |\psi \left( \overline{g}%
\right) \rangle =0.
\end{equation}%
We find this consequence stems from the non-degeneracy of the eigenstate of
the QRM. Actually we can prove more generally for a parameter $\lambda $ that the
vanishing relation
\begin{equation}
\langle \psi ^{\prime }\left( \lambda \right) |\psi \left( \lambda \right)
\rangle =0  \label{zero-dF-term}
\end{equation}%
holds for a general non-generate state $\psi \left( \lambda \right) $.
Indeed, we can always chose the general $\psi \left( \lambda \right) $ to be
real%
\begin{equation}
\psi ^{\ast }\left( \lambda ,x\right) =\psi \left( \lambda ,x\right)
\end{equation}%
at any position $x$, up to an irrelevant total phase, since otherwise
supposed linear-independent real part and imaginary part of a complex wave
function would yield degenerate eigenstates contradictorily to the
non-degenerate assumption. Thus, with the real wave function, we have
\begin{eqnarray}
\frac{d}{d\lambda }\langle \psi \left( \lambda \right) |\psi \left( \lambda
\right) \rangle  &=&\langle \psi ^{\prime }\left( \lambda \right) |\psi
\left( \lambda \right) \rangle +\langle \psi \left( \lambda \right) |\psi
^{\prime }\left( \lambda \right) \rangle   \nonumber \\
&=&2\langle \psi ^{\prime }\left( \lambda \right) |\psi \left( \lambda
\right) \rangle .  \label{dNorm}
\end{eqnarray}%
On the other hand, from the normalization condition $\langle \psi \left(
\lambda \right) |\psi \left( \lambda \right) \rangle =1$ we have $\frac{d}{%
d\lambda }\langle \psi \left( \lambda \right) |\psi \left( \lambda \right)
\rangle =0$, which leads to $\langle \psi ^{\prime }\left( \lambda \right)
|\psi \left( \lambda \right) \rangle =0$ in combination with Eq. (\ref{dNorm}%
). As a result, we can simplify the QFI with respect to a single parameter $%
\lambda $ as
\begin{equation}
F_{Q}=4\langle \psi ^{\prime }\left( \lambda \right) |\psi ^{\prime }\left(
\lambda \right) \rangle   \label{zero-dF-term-General}
\end{equation}%
for non-degenerate egein states.

\begin{figure*}[t]
\includegraphics[width=2\columnwidth]{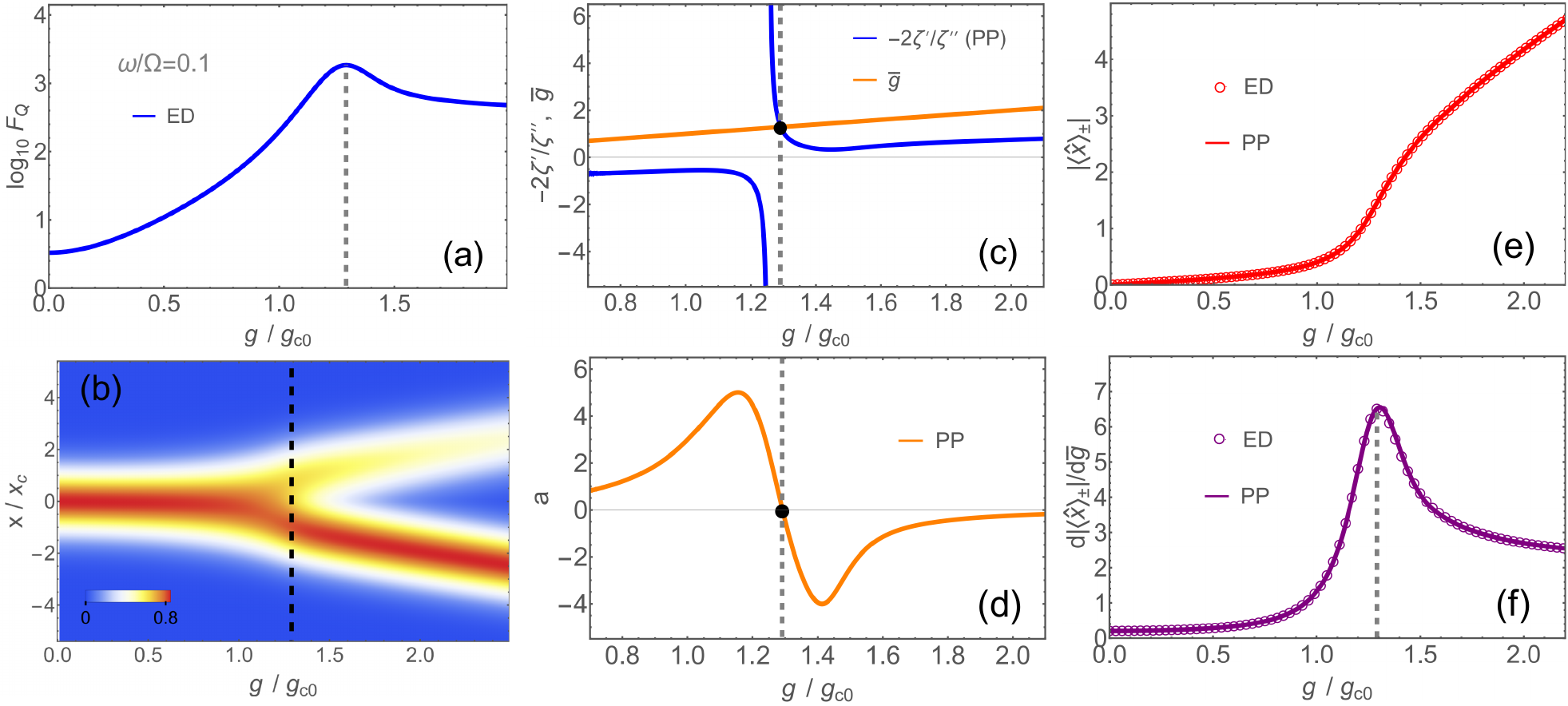}
\caption{Bridge of the QFI and transition property around the transition. (a) The transition position in $F_Q$ (solid line).
(b) The distribution probability in spin-up component.
(c) Transition coupling by crossing point (black dot) of $g$ (orange (light gray) solid line) and $-2\zeta^\prime/\zeta^{\prime\prime} $ in \eqref{gc-Acceleration}.
(d) Polaron acceleration $a$ versus $g$.
(e) The spin-component displacement expectation $\langle \widehat{x}\rangle_{\pm}$. The PP (solid line) yields a same result as the ED (circles).
(f) The spin-component displacement velocity $d\langle \widehat{x}\rangle_{\pm}/d\overline{g}$.
Here $\omega/\Omega=0.1$ and the vertical dashed lines in all panels mark the transition coupling at $F_Q^{\rm max}$. The $a=0$ point (black dot) in (d) indicates the maximum polaron velocity at transition similarly to the expectation in (f).
}
\label{fig-gc-max-velocity}
\end{figure*}

\subsection{Maximum effective velocity around transition}
\label{Sect-v-max}

The transition coupling identified by the QFI is then determined by the peak
position
\begin{equation}
\frac{dF_{Q}\left( \overline{g}\right) }{d\overline{g}}=4\frac{d\langle \psi
^{\prime }\left( \overline{g}\right) |\psi ^{\prime }\left( \overline{g}
\right) \rangle }{d\overline{g}}=0,  \label{dFq=0}
\end{equation}
where we have taken Eq. (\ref{zero-dF-term}) into account to use (\ref{zero-dF-term-General}).

In the polaron picture essentially the process of the
transition in the ground state of the QRM is the splitting or separating of the polarons under the competition of the tunneling energy and the potential energy. The tunneling energy between the two spin components is negative in the negative parity, while the potentials $v_{\sigma _z}(x)$ in \eqref{Hx} have different values for the two spin components as they are separated by the effective coupling $\widetilde{g}$. When the tunneling energy is dominant the polarons tend to stay around the origin $x=0$ to gain a maximum negative tunneling energy, otherwise when the potential cost is too high to stay around the origin the polarons leave the origin and the transition occurs. Such a picture of transition agrees with the ED result in Figure \ref{fig-gc-max-velocity}(b).

Note tunneling energy is proportional to the overlap of the poalrons, while the frequency renormalization can extend the polaron wavepackets to increase the overlap. Around the transition the polaron frequency nearly reaches the maximum
renormalization to keep the remnant tunneling energy as much as possible, with $\xi _{i}^{\prime }\left( \overline{g}\right) \approx 0$
around the transition~\cite{Ying2015}. On the other hand,
around and beyond the transition the overlap between the separated polarons becomes exponentially small. As an approximation
we can neglect the effect of weight variation in the polaron splitting due
to the insight that the weight lost in one polaron would goes to the other
polaron, which may not much affect the variation of the total contribution
of the wave function to the QFI. As anther simplification we assume the same same frequency renormalization $\xi$ and displacement renormalization magnitude $\zeta$ for the separating polarons. Under all these considerations, the QFI around the transition is simplified to be
\begin{equation}
F_{Q}\left( \overline{g}\right) \approx \frac{\Omega }{\omega }\left[ \zeta
^{\prime }\left( \overline{g}\right) \overline{g}+\zeta \left( \overline{g}\right) \right] ^{2}\xi \left( \overline{g}%
\right) \label{Fq-PP-at-gc}
\end{equation}
in the leading order.
Then from Eq. (\ref%
{dFq=0}) we find the transition coupling in the PP
\begin{equation}
\overline{g}_{c{\rm F}}^{{\rm PP}}=\frac{2\zeta ^{\prime }\left( \overline{g}%
\right) }{-\zeta ^{\prime \prime }\left( \overline{g}\right) }.
\label{gc-Acceleration}
\end{equation}
Here we have defined the derivatives $\zeta ^{\prime }\left( \overline{g}\right) =\frac{d}{d\overline{g}}%
\zeta \left( \overline{g}\right) $ and $\zeta ^{\prime \prime }\left(
\overline{g}\right) =\frac{d^{2}}{d\overline{g}^{2}}\zeta \left( \overline{g}%
\right) $.

The solution (\ref{gc-Acceleration}) can be re-arranged to be
\begin{equation}
\frac{d^{2}}{dg^{2}}\left( \zeta \widetilde{g}\right) =0.
\end{equation}%
Note $\widetilde{g}=\sqrt{2}g/\omega $ is the potential position and $x_p=\zeta
\widetilde{g}$ is the polaron displacement, thus it is also equivalent to%
\begin{equation}
a\equiv \frac{d^{2}}{dg^{2}}x_p=0.  \label{a=0}
\end{equation}%
If we vary $g$ with a uniform speed, $a$ is actually the effective
acceleration of the polaron in the increase of the coupling up to a square
factor of the increasing speed of the coupling. In such a sense, the
transition condition (\ref{a=0}) means a vanishing polaron acceleration or
the maximum polaron velocity (the first derivative $dx_p/dg$) with respect to
the increase of the coupling.

In turn, when we recall the relation \eqref{Fq-PP-at-gc}, the FQI around the transition is then endowed a more physical connotation
to be a renormalized effective kinetic energy
\begin{equation}
F_{Q}\left( \overline{g}\right) \approx \frac{1}{2}m_{F}\overline{v}_{p}^{2}
\end{equation}%
with renormalized mass, position and velocity $m_{F},\overline{x}_{p},%
\overline{v}_{p}$:
\begin{equation}
m_{F}=2\frac{\Omega }{\omega }\xi \left( \overline{g}\right) ,\quad
\overline{v}_{p}=\frac{d\overline{x}_{p}}{d\overline{g}},\quad \overline{x}%
_{p}=\zeta \overline{g}.
\end{equation}%

Figure \ref{fig-gc-max-velocity} illustrates an example at $\omega /\Omega
=0.1$ to show the validity of the relations (\ref{gc-Acceleration}) and (\ref%
{a=0}). Here in the figure panel (a) shows the QFI in logarithm numerically
calculated by the ED, with the vertical dashed line marking the transition
coupling $g_{c{\rm F}}$. Panel (b) shows the distribution probability by the
ED, one sees that the wave packet stays around the origin $x=0$ before the
transition while it splits into two wave packets and depart from each other
after the transition. The wave packets represent the polarons. Panel (c)
shows the crossing point (black dot) of the $\overline{g}$ line and the
curve of $2\zeta ^{\prime }\left( \overline{g}\right) /[-\zeta ^{\prime
\prime }\left( \overline{g}\right) ]$ in (\ref{gc-Acceleration}) by the main
polaron with a larger weight $\alpha $, which is the solution for $\overline{%
g}_{c{\rm F}}^{{\rm PP}}$. We see that $\overline{g}_{c{\rm F}}^{{\rm PP}}$
is in good agreement with the accurate $g_{c{\rm F}}$ (vertical dashed
line). Panel (d) tracks the corresponding evolution of the polaron
acceleration $a$, the zero point (black dot) also matches well with the
transition coupling $g_{c{\rm F}}$.

\begin{figure}[t]
\includegraphics[width=0.8\columnwidth]{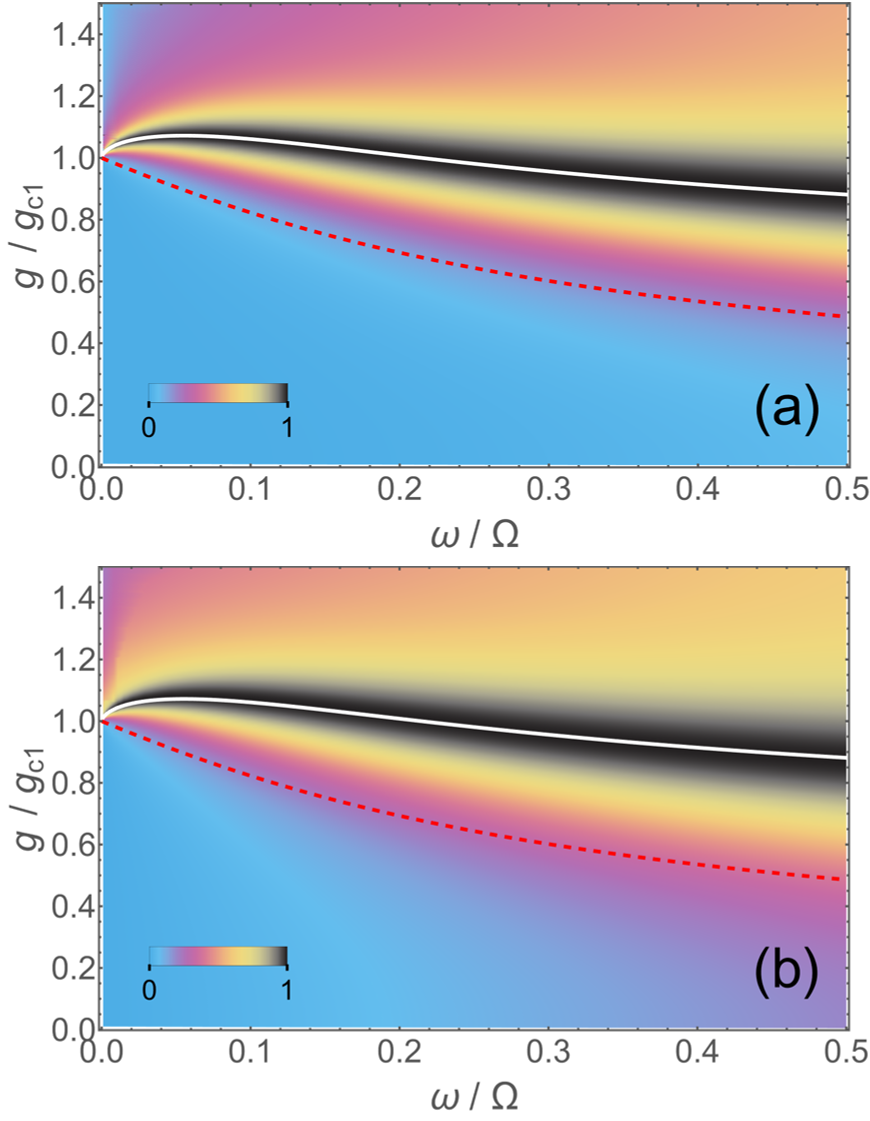}
\caption{Density plots of the QFI $F_Q$ (a) and the susceptibility of single-photon
absorption rate $d\left\vert \langle \widehat{x%
}\rangle _{\pm }\right\vert /dg$ (b), both renormalized by their maxima at fixed frequencies. Their maxima coincide with each other, which is well captured by $g_{c2}$ in Eq.\eqref{gc2} (white solid line) but missed by the conventional $g_{c0}$ (red dashed line).
}
\label{fig-Densityplot-Fisher-dxdg}
\end{figure}

\section{Bridging QFT to transition observable: Transition coincidence with maximum susceptibility of single-photon
absorption rate}

\label{Sect-dx-expectation}

The above discussion with the transition equation \eqref{a=0} obtained from the QFI inspires us to bridge the QFI to a transition observable.
The effective velocity and acceleration in Sec.\ref{Sect-v-max} is defined
on a polaron. We can also check the position expectation $\langle \widehat{x}%
\rangle _{\pm }$ for the spin components which is polaron-picture
independent and are available both in the ED and in the PP. In the PP we
have
\begin{equation}
\langle \widehat{x}\rangle _{+}=\alpha ^{2}\langle \varphi _{\alpha }|%
\widehat{x}|\varphi _{\alpha }\rangle +\beta ^{2}\langle \varphi _{\beta }|%
\widehat{x}|\varphi _{\beta }\rangle +2\alpha \beta \langle \varphi _{\alpha
}|\widehat{x}|\varphi _{\beta }\rangle
\end{equation}%
subject to the normalization condition $\alpha ^{2}+\beta ^{2}+2\alpha \beta
\langle \varphi _{\alpha }|\varphi _{\beta }\rangle =1$. The spin-down
component has the opposite sign, $\langle \widehat{x}\rangle _{-}=-\langle
\widehat{x}\rangle _{+}$, due to the parity symmetry. In the ED%
\begin{equation}
\langle \widehat{x}\rangle _{\pm }=\langle \psi _{\pm }|\widehat{x}|\psi
_{\pm }\rangle =\langle \psi _{\pm }|\frac{a^{\dagger }+a}{\sqrt{2}}|\psi
_{\pm }\rangle =\sqrt{2}\langle \psi _{\pm }|a|\psi _{\pm
}\rangle \label{absorption-rate}
\end{equation}%
is defined directly on the ED eigenstate $\left\vert \psi _{\pm
}\right\rangle $ expanded on the Fock states.

We plot the amplitude $%
\left\vert \langle \widehat{x}\rangle _{\pm }\right\vert $ in Fig.~\ref{fig-gc-max-velocity}(e,f)
where in panel (e) one sees that the results by the PP
(solid line) and the ED (circles) completely coincide with each other.
Similarly to the introduced velocity $dx/dg$ in the PP, in panel (f) we also
present the evolution of $d\left\vert \langle \widehat{x}\rangle _{\pm
}\right\vert /dg$ which, if getting out of the PP language, is actually the
susceptibility of the displacement or the single-photon absorption rate, as indicated by Eq.~\eqref{absorption-rate}, in
response to the coupling variation. Corresponding to the zero velocity
(maximum acceleration) in the PP, we see that $d\left\vert \langle \widehat{x%
}\rangle _{\pm }\right\vert /dg$ really reaches the maximum around the
transition (vertical dashed line).

Density plots in the $\omega$-$g$ plane for the QFI and the susceptibility of single-photon
absorption rate $d\left\vert \langle \widehat{x%
}\rangle _{\pm }\right\vert /dg$ are provided in Fig.~\ref{fig-Densityplot-Fisher-dxdg}. We see
that the maxima (black) of the QFI and $d\left\vert \langle \widehat{x%
}\rangle _{\pm }\right\vert /dg$ coincide with each other in the wide frequency regime,
as indicated by their simultaneous agreements with $g_{c2}$ in \eqref{gc2} (white solid line).
In contrast, the coinciding maxima are missed by the conventional transition coupling $g_{c0}$ (red dashed line).

Thus, via the transition we bridge the
QFI maximum point, the maximum polaron velocity and the maximum
susceptibility of the single-photon absorption rate.

\section{Conclusions}

\label{Sect-Conclusions}

We have combined the quantum Fisher information (QFI) and the variational
polaron picture (PP) to identify and extract the accurate transition
couplings for the quantum phase transition of the quantum Rabi model (QRM).
With the combined QFI-PP analysis we also gain some implication
and insight for the underlying physics of transition in the QRM.

The continuing interest on the QRM is only lying in the fact that it is a fundamental model for light-matter interactions
but also is attracted by the quantum phase transition it possesses which can be applied for the critical quantum
metrology. In quantum metrology, the square root of the QFI represents the precision bound of
experimental measurement. In the present work we have conversely used the
peak location of the QFI to identify the transition couplings in the QRM. By
the QFI result from exact diagonalization (ED) we have demonstrated that
transition couplings at finite frequencies much deviate from the
conventional one which is exact actually only right at zero frequency.
Inspired by the fractional-power-law behavior in the influence of polaron
frequency renormalization on the expression improvement for the transition
coupling, we have obtained an accurate expression of the transition coupling
which coincides with the numeric transition couplings by the QFI in the
variation of frequency.  Besides the implication acquired from the integer/fractional power
law analysis for a deeper understanding of the essence of transition, the transition coupling can provide an accurate
reference in the practical applications of the quantum phase transition in
the quantum metrology, since in practice the bosonic mode always has a
non-zero frequency which invalidates the frequency-dependence of the
conventional transition coupling.

We have also formulated the QFI in the PP. The PP is capable of analytically
reproducing the numeric QFI by ED, without a
heavy computation cost as in the ED. On the other hand, the QFI in the PP
can compensate for the missing data of the ED in the regime where the ED
cannot access due to the demanding need of large basis number. Such a
situation emerges in the large coupling regime, especially at low
frequencies where the phase transition is sharp and provides the best
sensitivity resource for raising the measurement precision. Therefore, the
formulation of the QFI in the PP prepares an accurate and convenient tool to
get the precision bound in covering the entire parameter regime, which may
also be helpful in applications of quantum metrology.

From the QFI in the PP we have also come across the vanishing of the
first-derivative term in the QFI in the QRM. We find this vanishing
consequence comes from the non-degeneracy of the eigenstate. The conclusion
has been extended for general non-degenerate pure states, which simplifies
the expression of the QFI.

Finally from the QFI in the PP we see that the transition occurs with a zero
polaron acceleration or maximum polaron velocity. Correspondingly the
susceptibility of single-photon absorption rate reaches the maximum around
the transition. This finding might add some insight in bridging the QFI and
the physical properties at the transition.

As a closing remark, it is worthwhile to mention that, besides the
transition in the QRM addressed in this work, similar transitions~\cite{LiuM2017PRL,Liu2021AQT,Ying-2021-AQT,Ying-Stark-top,Ying-gapped-top,Ying-2018-arxiv,Ying2020-nonlinear-bias,Ying2022-Metrology}
also occur at low frequencies in other forms of light-matter interactions,
such as the anisotropic coupling~\cite{PRX-Xie-Anistropy,LiuM2017PRL,Ying-2021-AQT,Ying-Stark-top,Ying-gapped-top,Ying-Spin-Winding,Yimin2018}, the mixture with nonlinear two-photon coupling~\cite{Ying-2018-arxiv,Ying2020-nonlinear-bias,Ying2022-Metrology} and
the Stark nonlinear coupling~\cite{Eckle-2017JPA,Maciejewski-Stark,Ying-Stark-top,Xie2019-Stark,Ying-JC-winding}. On the other hand, although the two-polaron description in the PP has
reached a very high accuracy, extension to the multi-polaron representation%
\cite{CongLei2017} is direct if even higher accuracy is needed. Our
treatment addressed in the present work concerning the QFI and the
transition coupling can be readily applied to these systems, which we shall
leave for some future works.

\section*{Acknowledgements}

This work was supported by the National Natural Science Foundation of China
(Grants No.~11974151 and No.~12247101).

\appendix

\bigskip

\section{Inspiration for the transition coupling from frequency renormalization
effect of polarons}

\begin{figure}[t]
\includegraphics[width=0.8\columnwidth]{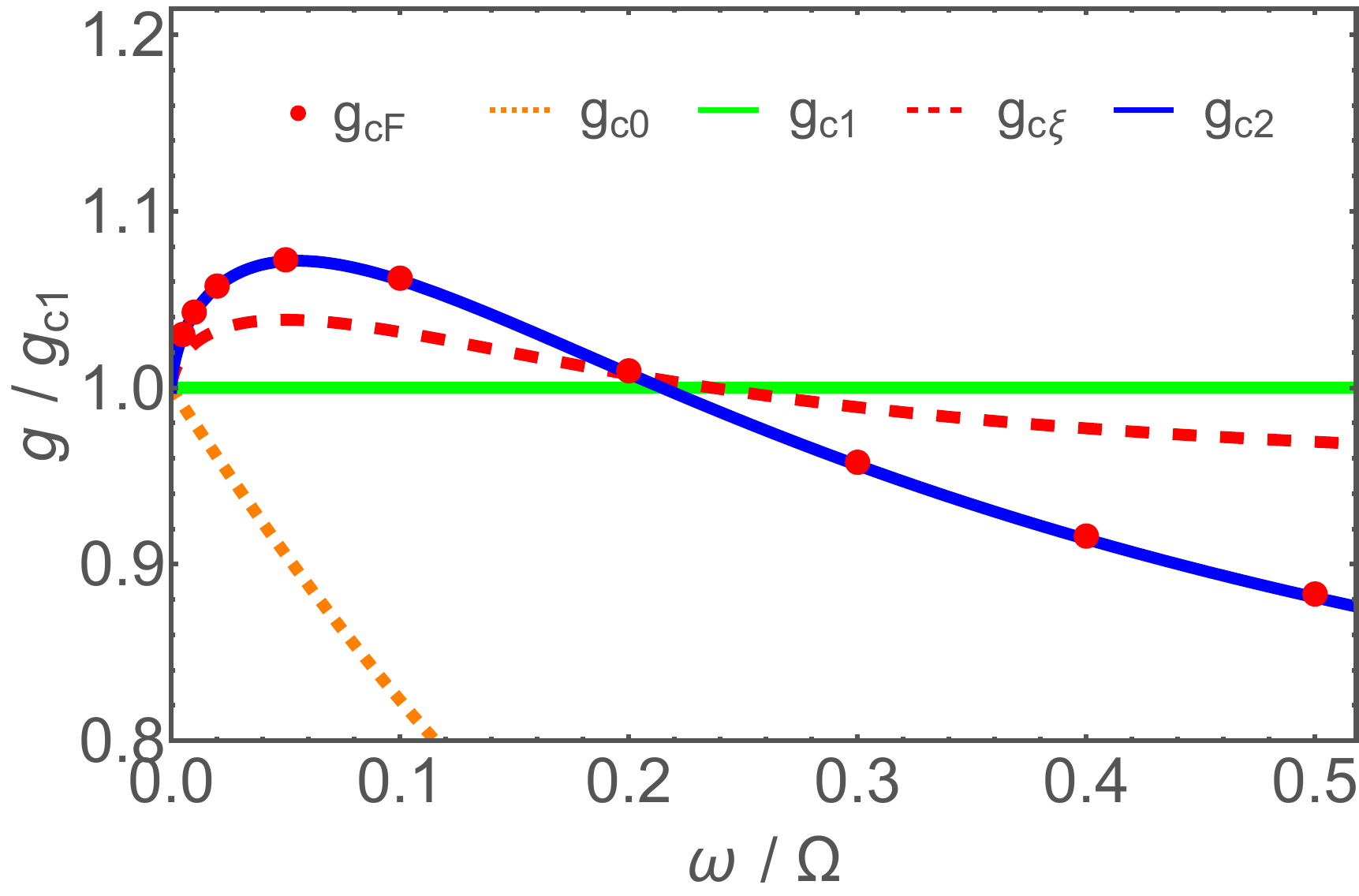}
\caption{Inspiration from the transition coupling $g_{c\xi}$ by including frequency renormalization. $g_{c\xi}$ (red dashed line) in \eqref{gC-by-squeezing} has some improvements over $g_{c0}$ (orange dashed) and $g_{c1}$ (green (light gray) solid line). Although $g_{c\xi}$ is still not accurate in comparison with $g_{cF}$ (red dots), it has the right direction of improvement and provides inspiration for formulation of $g_{c2}$ (blue (dark gray) solid line).
}
\label{fig-gc-squeezing}
\end{figure}

\label{Appendix-gC-ksi}

From Fig. \ref{fig-gc-max-velocity} in the main text, one may realize that
the transition is essentially the splitting of the wave packet, which can be
well described by the separation of the two polaron represented by $\varphi
_{i}\left( \overline{g},x\right) =\left( \xi _{i}/\pi \right) ^{1/4}\exp %
\left[ -\xi _{i}\left( x+x_{i}\right) ^{2}/2\right] $ in (\ref%
{polaron-Wave-Function}). Here $x_{\alpha }=\zeta _{\alpha }\widetilde{g}$
and $x_{\alpha }=\zeta _{\beta }\widetilde{g}$ with $\widetilde{g}=\sqrt{2}%
g/\omega .$ As a simplification we can assume two same polarons by setting
\begin{equation}
\zeta _{\alpha }=-\zeta _{\beta }=\zeta ,\quad \xi _{\alpha }=\xi _{\beta
}=\xi .  \label{Same-polarons}
\end{equation}
Here the minus sign of $\zeta _{\beta }$ denotes the displacement direction
opposite to the other polaron $\alpha $. Thus the distance of the two
polarons is $2\zeta \widetilde{g}$. The overlap of the two polarons is
decreasing exponentially in the separation. The overlapping degree is
decided by the crossing point of the two polaron wave packets which is
moving away from the peak by a distance $d$. One can regard the separation
to be basically complete when $d$ reaches some point $d_{c}$ where the
polaron overlap becomes small enough to complete the transition. The
distance relation around the transition is then given by%
\begin{equation}
2\zeta \widetilde{g}=2d_{c}.  \label{d=dc}
\end{equation}%
Note that in the low-frequency limit we have the explicit displacement
renormalization factor \cite{Ashhab2013,Ying2015,Ying-2018-arxiv}
\begin{equation}
\zeta \approx \sqrt{1-\frac{g_{c0}^{4}}{g^{4}}}  \label{zeta}
\end{equation}%
starting from the transition, which can be applied to obtain improved expressions of the transition coupling.

\subsection{$g_{c0}$ by classical mass point}

To mark the difference from the polaron frequency renormalization we set $%
d_{c}=d_{c1}/\sqrt{\xi }$ where $d_{c1}$ is the transition distance in the
absence of frequency renormalization. If the polaron wave packet is regarded
as a classical mass point, then the separation process is immediately complete at
$d_{c}=0$, substitution of which in (\ref{d=dc}) retrieves the conventional
transition coupling%
\begin{equation}
g_{c}=g_{c0}=\frac{\sqrt{\omega \Omega }}{2}
\end{equation}
as in \eqref{gc0}.

\subsection{$g_{c1}$ by neglecting polaron frequency renormalization}

However, in quantum mechanics the mass point should be replaced by the wave
packet and $d_{c}$ is finite. As a first order of improvement with finite $%
d_{c}$, we can neglect the polaron frequency renormalization by setting $\xi
=1$. We can judge the transition by a point where the wave function $%
\varphi _{i}$ at the crossing point decays to an exponentially small value by the
ratio $\varphi _{i}/\varphi _{i}^{{\rm peak}}=e^{-2}$ which occurs at $%
d_{c}=2$. Then Eq. (\ref{d=dc}) becomes
\begin{equation}
\sqrt{1-\frac{g_{c0}^{4}}{g_{c}^{4}}}\frac{\sqrt{2}g_{c}}{\omega }=2
\end{equation}%
which yields
\begin{equation}
g_{c}=g_{c1}=\sqrt{\omega ^{2}+\sqrt{\omega ^{4}+g_{c0}^{4}}}  \label{gc=gc1}
\end{equation}%
in (\ref{gc1}) as an improved transition coupling over $g_{c0}$.

\subsection{$g_{c\protect\xi }$ by including polaron frequency
renormalization}

We now pick up the polaron frequency renormalization and notice the
frequency-displacement scaling relation $\xi /\zeta \approx 1$ starting from
the transition in the low frequency limit, i.e. \cite{Ying2015}%
\begin{equation}
\xi \approx \zeta \approx \sqrt{1-\frac{g_{c0}^{4}}{g^{4}}}.  \label{Zeta-Ksi}
\end{equation}%
In such a consideration Eq. (\ref{d=dc}) becomes much more nonlinear
\begin{equation}
\sqrt{1-\frac{g_{c0}^{4}}{g_{c}^{4}}}\frac{\sqrt{2}g_{c}}{\omega }=\frac{%
d_{c1}}{\sqrt{\xi }}=\frac{d_{c1}}{\left( 1-\frac{g_{c0}^{4}}{g_{c}^{4}}%
\right) ^{1/4}}.  \label{gc-eqaution-ksi-polaron}
\end{equation}%
Nevertheless, Equation (\ref{gc-eqaution-ksi-polaron}) still can be
analytically solved, with the explicit solution being%
\begin{equation}
g_{c}=g_{c\xi }=\sqrt[4]{g_{c0}^{4}+\frac{\omega _{c1}^{4}}{12}\left(
f^{4}+1+f^{-4}+24f^{-4}\tilde{g}_{c0}^{4}\right) },\label{gC-by-squeezing}
\end{equation}%
where we have defined $\omega _{c1}=d_{c1}\omega ,\quad \tilde{g}_{c0}=\frac{%
g_{c0}}{\omega _{c1}}$, and
\begin{equation}
f=\sqrt[12]{1+36\tilde{g}_{c0}^{4}+216\tilde{g}_{c0}^{8}+24\sqrt{3}\tilde{g}%
_{c0}^{6}\sqrt{27\tilde{g}_{c0}^{4}+1}}.
\end{equation}

\subsection{Inspiration of fractional-power-law expansion for $g_{c2}$}

\label{Appendix-fractionl-law}

In Fig. \ref{fig-gc-squeezing} we give a comparison for $g_{c0}$ (dotted
line), $g_{c1}$ (green (light-gray) solid line) and $g_{c\xi }$ (red dashed
line) with $d_{c1}=1.9$. Compared with $g_{c{\rm F}}$ (dots), $g_{c1}$ is
qualitatively better but quantitatively still not accurate enough, as also
mentioned in the main text. By taking into the effect of the polaron
frequency renormalization, $g_{c\xi }$ yields some more improvements over $g_{c1}$, especially in the low-frequency tendency.
Although $g_{c\xi }$ does not achieve a perfect accuracy due to the
simplifications we have introduced in the above discussion, e.g. in (\ref%
{Same-polarons}) and (\ref{Zeta-Ksi}), the improvements added by the polaron
frequency renormalization is in the correct direction. Note here the
frequency renormalization effect manifests itself a behavior
of fractional-power-series law in low frequencies
\begin{eqnarray}
g_{c\xi } &=&g_{c0}\left[ 1+(\frac{d_{c1}}{2})^{\frac{4}{3}}(\frac{\omega }{%
\Omega })^{\frac{2}{3}}+\frac{7}{6}(\frac{d_{c1}}{2})^{\frac{8}{3}}(\frac{%
\omega }{\Omega })^{\frac{4}{3}}+\cdots \right]   \nonumber \\
&=&g_{c0}\left[ 1+\sum_{n=1}^{\infty }c_{n}^{\xi }\left( \frac{\omega }{%
\Omega }\right) ^{n\frac{2}{3}}\right] ,  \label{gC-kesi-fract-power-law}
\end{eqnarray}%
which inspires us to propose the expressions $g_{c2}$ and $g_{c2}^{{\rm %
fitting}}$ in (\ref{gc2}) and (\ref{gc-fitting}) for the transition
coupling. It turns out that $g_{c2}$ and $g_{c2}^{{\rm fitting}}$ are very
accurate with only a couple of terms in the fractional power law expansion.

\begin{figure}[t]
\includegraphics[width=1\columnwidth]{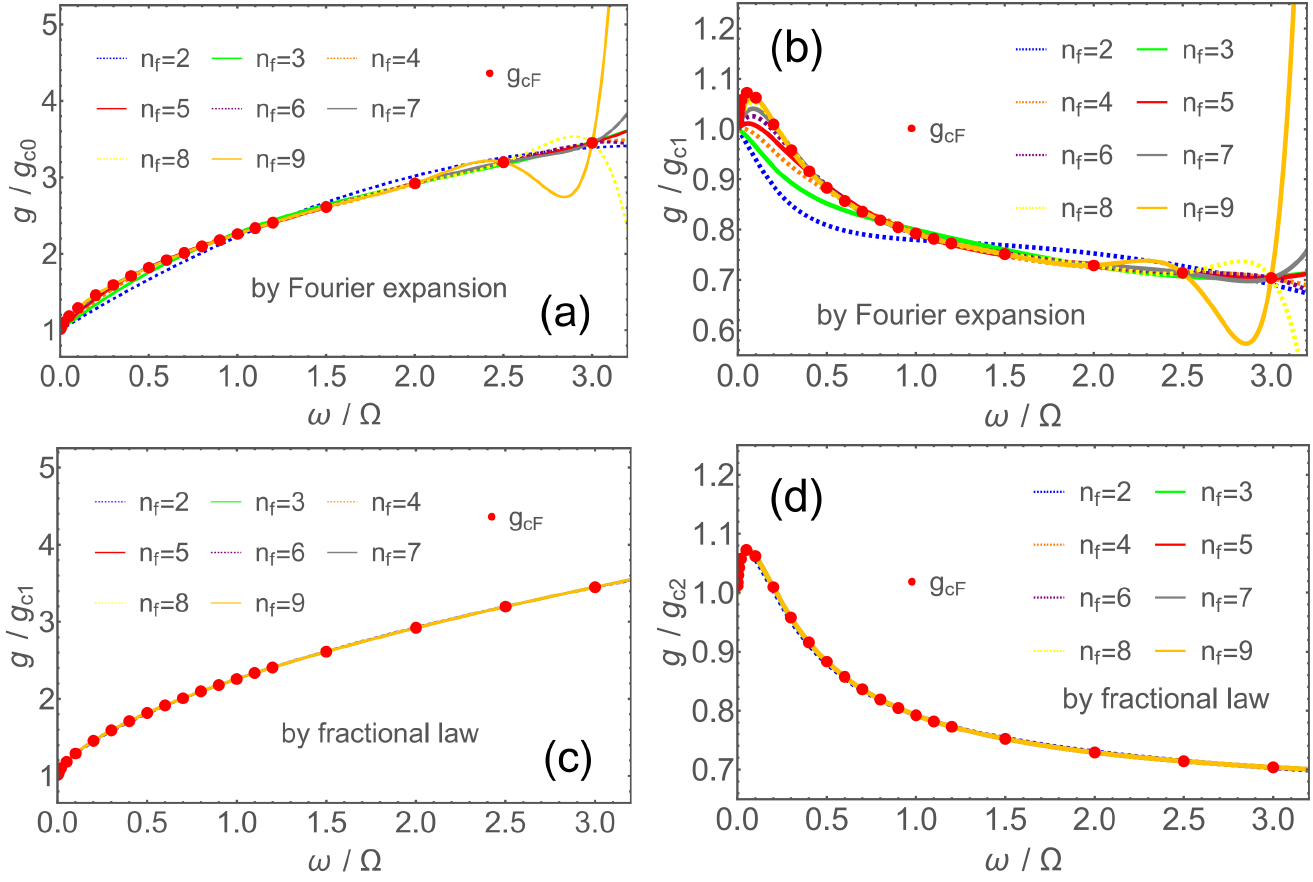}
\caption{Comparison of fitting of the transition coupling $g_{c{\rm F}}$ by Fourier expansion (a,b) and by fractional-power law (c,d) with different expansion order $n_f$. (a,c) $g$ scaled by $g_{c0}$ (b,d) $g$ scaled by $g_{c1}$.
}
\label{fig-Fourier-fitting}
\end{figure}

\subsection{Unfavorable mathematical expansion by Fourier Series}

\label{Appendix-Fourier-Expansion}

It should be stressed that accuracy and efficiency of the fractional-power
form of $g_{c2}$ and $g_{c2}^{{\rm fitting}}$ in (\ref{gc2}) and (\ref%
{gc-fitting}) are based on the physical analysis from frequency
renormalization effect of polarons, as addressed in last sections. One might
speculate that a mathematical expansion by Fourier Series%
\begin{equation}
g_{c}^{{\rm Fourier}}=g_{c0}\left[ 1+\sum_{n=1}^{n_{f}}c_{n}^{{\rm Fourier}%
}\left( \frac{\omega }{\Omega }\right) ^{n}\right]
\end{equation}%
would also work without need of the fractional-power form. It is true that a Fourier
expansion can mathematically make a fitting finally, however such a
Fourier-series fitting is much more inefficient, without mentioning that it
does not provide any physical insight. We show some orders of the
Fourier-expansion fitting in Fig. \ref{fig-Fourier-fitting} via the
least-squares fit. One sees in panel (a) by the coupling scale $g_{c0}$ that
$n_{f}=2$ (blue dotted most-deviated line) in Fourier expansion does not fit
well at all, the qualitative deviations can be seen more clearly in panel (b) by
the coupling scale $g_{c1}$. Higher orders up to $n_{f}=9$ approach to
some convergence at low frequencies but large oscillations appear at high
frequencies due to sparser reference data of $g_{c{\rm F}}$ (red dots). In a
sharp contrast, fitting by the fractional law in (\ref%
{gC-kesi-fract-power-law}) already reaches a good convergence even at $%
n_{f}=2$. Moreover, larger $n_{f}$ converge well without the oscillation
problem as the Fourier expansion even in the same sparse reference data.

In fact, the expression (\ref{gc=gc1}) obtained by neglecting polaron frequency
renormalization, if expanded, has an integer-power-series law as $g_{c}^{%
{\rm Fourier}}$
\begin{eqnarray}
g_{c1} &=&g_{c0}\left( 1+\frac{2\omega }{\Omega }+\frac{2\omega ^{2}}{\Omega
^{2}}-\frac{4\omega ^{3}}{\Omega ^{3}}-\frac{10\omega ^{4}}{\Omega ^{4}}%
+\cdots \right)   \nonumber \\
&=&g_{c0}\left[ 1+\sum_{n=1}^{\infty }c_{n1}\left( \frac{\omega }{\Omega }%
\right) ^{n}\right] .  \label{gc1-Expansion}
\end{eqnarray}%
Both the inefficiency of fitting convergence and the coincidence of
integer-power-series law with $g_{c1}$ indicate that the Fourier-series
fitting $g_{c}^{{\rm Fourier}}$ does not capture the physical essence of
frequency renormalization effect around the transition in the QRM.

\begin{figure}[t]
\includegraphics[width=0.8\columnwidth]{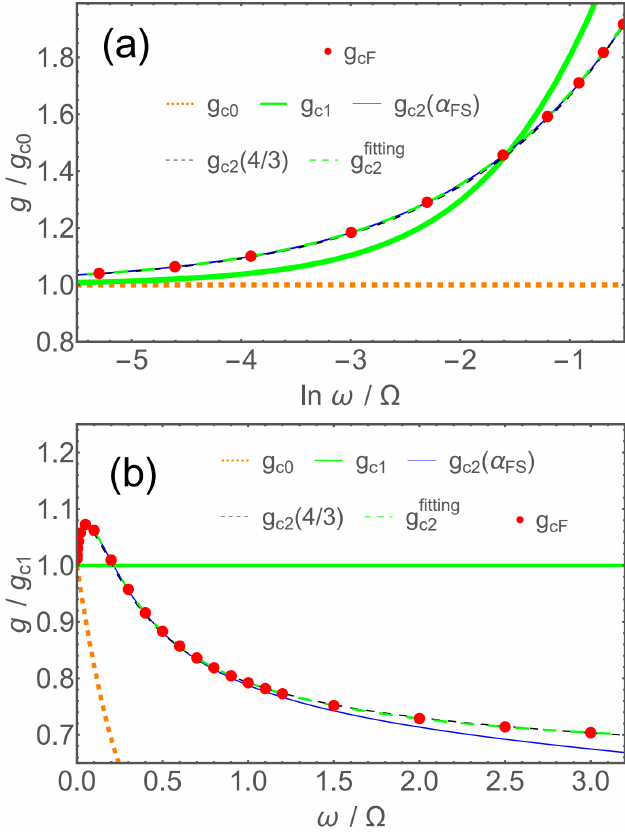}
\caption{Comparison of $g_{c2}$ with different coefficients in low-frequency regime (a) and high-frequency regime (b): $g_{c2}$ in Eq. \eqref{gc2} with $\alpha _{\rm FS}=1/137$  (thin blue dashed line), $g_{c2}$ in Eq. \eqref{gc2-34} with $c_1=4/3$  (thin black dashed line) and $g_{c2}^{\rm fitting}$ in \eqref{gc-fitting} with $c_1=1.3715$, $c_2=0.1311$, $c_3=0.0184$ (green long-dashed line). The result of $g_{c{\rm F}}$ (red dots) is the accuracy reference. $g_{c0}$ (thick orange dotted line) and $g_{c1}$ (thick green solid line) are taken as the coupling scale in (a) and (b).
}
\label{fig-Full-winding}
\end{figure}

\section{Comparison of $g_{c2}$ in different coefficient choices}
\label{Appendix-compare-gc2}

In Sec. \ref{Sect-gc2-accurate} we provide different expressions of the
transition coupling (\ref{gc2})-(\ref{gc-fitting}). Here we make a
comparison for their difference. Figure \ref{fig-Full-winding} shows $g_{c2}$
in Eq. \eqref{gc2} with $\alpha _{{\rm FS}}=1/137$ (thin blue dashed line), $%
g_{c2}$ in Eq. \eqref{gc2-34} with $c_{1}=4/3$ (thin black dashed line) and $%
g_{c2}^{{\rm fitting}}$ in \eqref{gc-fitting} with $c_{1}=1.3715$, $%
c_{2}=0.1311$, $c_{3}=0.0184$ (green long-dashed line) in comparison with
the $g_{c{\rm F}}$ data (red dots) numerically extracted from the peak
locations of the QFI. As shown in Fig. \ref{fig-Full-winding}(a), the
expression \eqref{gc2} with $\alpha _{{\rm FS}}=1/137$ is very accurate and
is actually producing $g_{c{\rm F}}$ in the low frequency regime $\omega
/\Omega <0.5$. Although considerable deviations appear in the high frequency
regime $1<\omega /\Omega <3$ in Fig. \ref{fig-Full-winding}(b), the
transition loses the sense there as it becomes much broadened. Nevertheless,
the QFI always has a peak value, an expression valid for the entire
parameter regime would be a perfect goal. In this need, the expression %
\eqref{gc2-34} with $c_{1}=4/3$ is accurate up to $\omega /\Omega =3$,
despite it is slightly less accurate in the low frequency regime with an
error of the order $10^{-3}$ larger than \eqref{gc2}. Finally $g_{c2}^{{\rm %
fitting}}$ is very accurate really in the entire frequency regime up to $%
\omega /\Omega =3.$ Note $g_{c2}$ in \eqref{gc2} and \eqref{gc2-34} is in
the second order of $(\omega /\Omega )^{2n/3}$ power law, while $g_{c2}^{%
{\rm fitting}}$ is in the third order, which already reaches the convergence
as shown by Fig. \ref{fig-Fourier-fitting}(c,d) in Appendix \ref%
{Appendix-Fourier-Expansion}.

\section{Explicit expressions for the QFI in the variational
polaron picture}
\label{Appendix-Fq-PP-expressions}

Explicitly, by the wave function of the polaron in Eq. \eqref{polaron-Wave-Function}
for the non-derivative terms we have
\begin{eqnarray}
\langle \varphi _{\alpha }|\varphi _{\alpha }\rangle  &=&\langle \varphi
_{\beta }|\varphi _{\beta }\rangle =1,\quad  \\
\langle \varphi _{\alpha }|\varphi _{\beta }\rangle  &=&\frac{\sqrt{2}\xi
_{\alpha }^{1/4}\xi _{\beta }^{1/4}}{\sqrt{\xi _{\alpha }+\xi _{\beta }}}%
\exp [-\frac{\left( x_{\alpha }-x_{\beta }\right) ^{2}\xi _{\alpha }\xi
_{\beta }}{2\left( \xi _{\alpha }+\xi _{\beta }\right) }].
\end{eqnarray}
For the polaron-derivative terms in \eqref{1st-derivative-terms} we get
\begin{eqnarray}
\langle \frac{\partial \varphi _{i}}{\partial x_{i}}|\varphi _{j}\rangle  &=&%
\frac{\sqrt{2}\xi _{i}^{5/4}\xi _{j}^{5/4}\left( x_{j}-x_{i}\right) }{\left(
\xi _{i}+\xi _{j}\right) ^{3/2}f_{E}},
\\
\langle \varphi _{j}|\frac{%
\partial \varphi _{j}}{\partial x_{j}}\rangle &=&\frac{\sqrt{2}\xi
_{i}^{5/4}\xi _{j}^{5/4}\left( x_{i}-x_{j}\right) }{\left( \xi _{i}+\xi
_{j}\right) ^{3/2}f_{E}}, \\
\langle \frac{\partial \varphi _{i}}{\partial \xi _{i}}|\varphi _{j}\rangle
&=&\frac{\sqrt{2}\xi _{j}^{1/4}\{\xi _{i}^{2}+\xi _{j}^{2}[2\xi _{i}\left(
x_{i}-x_{j}\right) ^{2}-1]\}}{\xi _{i}^{3/4}\left( \xi _{i}+\xi _{j}\right)
^{5/2}f_{E}}, \\
\langle \varphi _{i}|\frac{\partial \varphi _{j}}{\partial \xi _{j}}\rangle
&=&\frac{\sqrt{2}\xi _{i}^{1/4}\{\xi _{j}^{2}+\xi _{i}^{2}[2\xi _{j}\left(
x_{i}-x_{j}\right) ^{2}-1]\}}{\xi _{j}^{3/4}\left( \xi _{i}+\xi _{j}\right)
^{5/2}f_{E}},
\end{eqnarray}
and in \eqref{2nd-derivative-terms} the explicit expressions read
\begin{widetext}
\begin{eqnarray}
\langle \frac{\partial \varphi _{i}}{\partial x_{i}}|\frac{\partial \varphi
_{j}}{\partial x_{j}}\rangle  &=&\frac{\sqrt{2}\{\xi _{i}+\xi _{j}-\xi
_{i}\xi _{j}\left( x_{i}-x_{j}\right) ^{2}\}}{\left( \xi _{i}+\xi
_{j}\right) ^{5/2}\xi _{i}^{-5/4}\xi _{j}^{-5/4}f_{E}}, \\
\langle \frac{\partial \varphi _{i}}{\partial x_{i}}|\frac{\partial \varphi
_{j}}{\partial \xi _{j}}\rangle  &=&\frac{\{\xi _{j}^{2}+\xi _{i}^{2}[2\xi
_{j}\left( x_{i}-x_{j}\right) ^{2}-5]-4\xi _{i}\xi _{j}\}}{2\sqrt{2}\left(
\xi _{i}+\xi _{j}\right) ^{7/2}\xi _{i}^{-5/4}\xi _{j}^{-1/4}\left(
x_{i}-x_{j}\right) ^{-1}f_{E}}, \\
\langle \frac{\partial \varphi _{i}}{\partial \xi _{i}}|\frac{\partial
\varphi _{j}}{\partial x_{j}}\rangle  &=&\frac{\{\xi _{i}^{2}+\xi
_{j}^{2}[2\xi _{i}\left( x_{i}-x_{j}\right) ^{2}-5]-4\xi _{i}\xi _{j}\}}{2%
\sqrt{2}\left( \xi _{i}+\xi _{j}\right) ^{7/2}\xi _{j}^{-5/4}\xi
_{i}^{-1/4}\left( x_{j}-x_{i}\right) ^{-1}f_{E}}, \\
\langle \frac{\partial \varphi _{i}}{\partial \xi _{i}}|\frac{\partial
\varphi _{j}}{\partial \xi _{j}}\rangle  &=&\frac{4\xi _{i}^{3}\xi
_{j}^{3}x_{ij}^{4}+\xi _{ij}^{+}(2\xi _{i}\xi _{j}x_{ij}^{2}-\xi
_{ij}^{+})(\xi _{i}^{2}+\xi _{j}^{2}-10\xi _{i}\xi _{j})}{8\sqrt{2}\left(
\xi _{i}+\xi _{j}\right) ^{9/2}\xi _{i}^{3/4}\xi _{j}^{3/4}f_{E}},
\end{eqnarray}
\end{widetext}
where we have defined $\xi _{ij}^{+}=\left( \xi _{i}+\xi _{j}\right) $, $%
x_{ij}=\left( x_{i}-x_{j}\right) $ and
$f_{E}=\exp \{\left(
x_{i}-x_{j}\right) ^{2}\xi _{i}\xi _{j}/[2\left( \xi _{i}+\xi _{j}\right) ]\}
$.
The dispalcement-derivative factor is
\begin{equation}
\frac{dx_{i}}{d\overline{g}}=\frac{d\zeta _{i}}{d\overline{g}}\overline{g}%
\sqrt{\frac{\Omega }{2\omega }}+\zeta _{i}\sqrt{\frac{\Omega }{2\omega }},
\end{equation}
explicitly.

The variational parameters \{$w _{i},\zeta _{i},\xi _{j}$\} for the ground state are determined by
minimization of the energy subject to normalization condition (\ref%
{normalization-condition}) of the wave function \cite{Ying2015}.

\end{document}